\documentclass[conference]{IEEEtran}
\IEEEoverridecommandlockouts
\usepackage{cite}
\usepackage{amsmath,amssymb,amsfonts}
\usepackage{dsfont}
\usepackage{algorithmic}
\usepackage{amsthm}
\usepackage{graphicx}
\usepackage{textcomp}
\usepackage{xcolor}
\usepackage{tabularx}
\usepackage[export]{adjustbox}
\usepackage{subcaption}
\usepackage{soul}
\usepackage{amsthm}

\def\BibTeX{{\rm B\kern-.05em{\sc i\kern-.025em b}\kern-.08em
    T\kern-.1667em\lower.7ex\hbox{E}\kern-.125emX}}

\newtheorem{lem}{Lemma}
\theoremstyle{definition}

\newcommand{\Hyp}[1]{\ensuremath{\mathbb{H}_{#1}}}

\newcommand{\testH}{\ensuremath{\underset{\Hyp{0}}{\overset{\Hyp{1}}{\gtrless}}}}

\newcommand{\SY}[1]{\textcolor{orange}{#1}}
\newcommand{\supprime}[1]{}
\begin{document}

\title{Neyman-Pearson Detector for Ambient Backscatter Zero-Energy-Devices Beacons \\
}

\author{ Shanglin Yang*\textsuperscript{o}, Jean-Marie Gorce\textsuperscript{o}, Muhammad Jehangir Khan\textsuperscript{o}, Dinh-Thuy Phan-Huy*, Guillaume Villemaud\textsuperscript{o} \\

*Orange Innovation/Networks, Châtillon, France \{name.surname\}@orange.com\\

\textsuperscript{o} INSA Lyon, Inria, CITI, EA3720, Villeurbanne, France \{name.surname\}@insa-lyon.fr

}

\maketitle

\begin{abstract}
Recently, a novel ultra-low power indoor wireless positioning system has been proposed. In this system, Zero-Energy-Devices (ZED) beacons are deployed in Indoor environments, and located on a map with unique broadcast identifiers. They harvest ambient energy to power themselves and backscatter ambient waves from cellular networks to send their identifiers. This paper presents a novel detection method for ZEDs in ambient backscatter systems, with an emphasis on performance evaluation through experimental setups and simulations. We introduce a Neyman-Pearson detection framework, which leverages a predefined false alarm probability to determine the optimal detection threshold. This method, applied to the analysis of backscatter signals in a controlled testbed environment, incorporates the use of BC sequences to enhance signal detection accuracy. The experimental setup, conducted on the FIT/CorteXlab testbed, employs a two-node configuration for signal transmission and reception. Key performance metrics, which is the peak-to-lobe ratio, is evaluated, confirming the effectiveness of the proposed detection model. The results demonstrate a detection system that effectively handles varying noise levels and identifies ZEDs with high reliability. The simulation results show the robustness of the model, highlighting its capacity to achieve desired detection performance even with stringent false alarm thresholds. This work paves the way for robust ZED detection in real-world scenarios, contributing to the advancement of wireless communication technologies.
\end{abstract}


\section{Introduction}
The enhancement of indoor self-localization for smartphones (SP) remains a critical topic that requires new technologies to progress, and effective detection mechanisms are also essential. Traditional Global Positioning System (GPS) based localization excels in outdoor environments \cite{b1}, prompting extensive exploration and commercialization of alternatives grounded in terrestrial wireless networks. Previous implementations, including technologies like 5G base stations and Wi-Fi access points, are providing accuracy to within a few meters \cite{Hi-Loc,Fast-Loc,DL-Loc,BT-Loc,BLE-Loc}. However, these methods are often limited by their high costs and energy demands. The sixth generation (6G) roadmap as proposed in the Hexa-X I European Flagship Project introduces an innovative approach with Crowd-Detectable zero-energy-devices (CD-ZED), which uses ambient backscatter for communication, promising a more sustainable localization service \cite{6GHexa-X,InnovHexa-X,AmB-ZED-DT}. Notably, these devices are under consideration by the 3rd Generation Partnership Project (3GPP) for potential integration into future mobile network standards, falling under the overarching concept of "Ambient Internet-of-Things" (A-IoT) \cite{3gpp_tr_38_848}. In \cite{AmB-ZED-DT,AmB-FSK-Alto,AmB-DemoLTE-Alto,ZED-6G-Papis}, a proof of concept of a ZED was proposed and we evaluated in \cite{ZED-Loc-SY} the coverage statistics of such ZED in simulated environments.

In this paper, a novel detector for the ZED based localization system studied in \cite{ZED-Loc-SY} is proposed. We use the Neyman-Pearson theory \cite{neyman1933} to optimize the detector. We do not change the ZED technology and coded sequences but replace the received dual frequency bandpass filters by dual correlators. This allows to rigorously determine the optimal Bayesian detector and to control the false alarm probability, which has not been taken into account in the initial design, despite its strong importance. 
The theory herein proposed exploits the Neyman Pearson formalism, and is highly relevant for engineering design, allowing to rely the detection performance as a function of the synchronization sequence length, and coverage requirements. Compared to previous works \cite{AmB-ZED-DT,AmB-FSK-Alto,AmB-DemoLTE-Alto,ZED-6G-Papis}, this is the first paper proposing a formal evaluation of the ZED performance.

\begin{figure}[htbp]
\centerline{\includegraphics[scale=0.65]{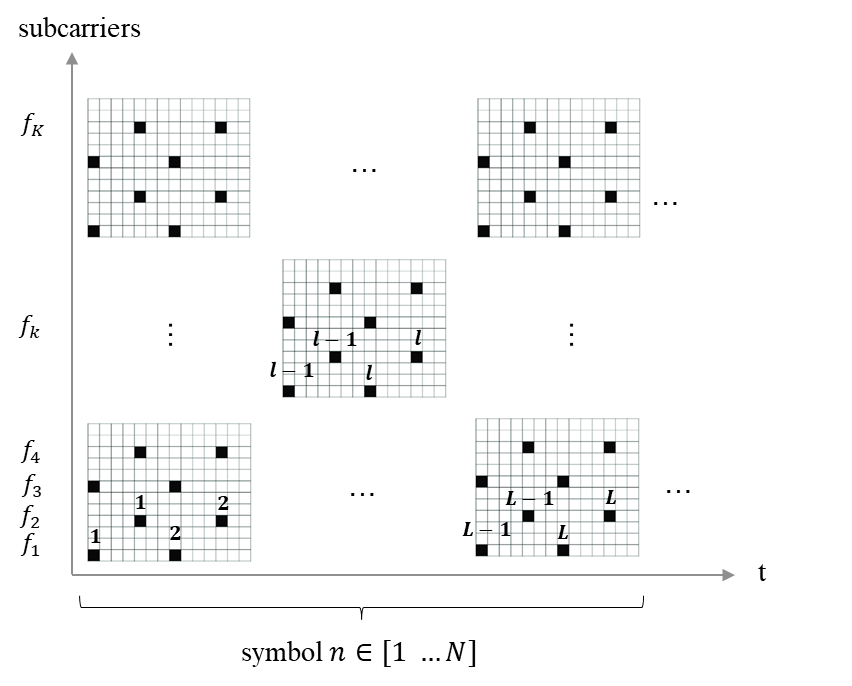}}
\caption{Pilots of 4G reference signal}
\label{fig:LTE}
\end{figure}

\section{System model}
In this section, we introduce the proposed system. The main changes with respect to \cite{AmB-ZED-DT,ZED-Loc-SY} are summarized.
This paper proposes a non-coherent receiver, motivated by experimental results showing the limitations of conventional receivers in maintaining phase coherence across OFDMA slots. Our observations indicate that traditional systems cannot reliably preserve phase over extended intervals. 

\subsection{ZED equivalent transmitter}
\begin{figure}[htbp]
\centerline{\includegraphics[scale=0.4]{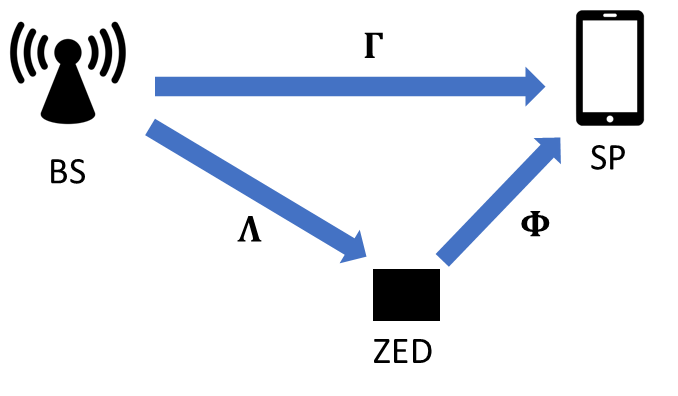}}
\caption{ZED based communication channel model.}
\label{fig:ZED}
\end{figure}

The communication between a ZED and an SP utilizes ambient 4G base station signals with reference signals (RS) located in specific resource elements of the OFDM frame (see Fig.\ref{fig:LTE}). Let $m(k,t(k,l)):([1,K],[1,L])\in \mathbb{R^+}$ noted $m(k,l)$ be the mapping function between carrier frequency index $k\in(1,K)$ and RS index $l\in(1,L)$ and time $t\in \mathbb{R^+}$, this function is illustrated in Fig.1. RS are distributed across $K=4N_{RB}$ subcarriers in the 4G standard. Each Transmission Time Interval (TTI) consists of 14 OFDM symbols, with $TTI = 14T^{ofdm}$ \cite{3gpp_ts_36_211}. For simplicity, we assume all RS have a constant magnitude of $ \sqrt{P_u} $, where $P_u$ is the pilot transmit power.
The ZED is a device that switches between passive and reflective states \cite{AmB-FSK-Alto,AmB-DemoLTE-Alto}. In passive mode, it has no effect on the BS waves, while in reflective mode, it reflects the signal, creating an alternative path that slightly alters the channel response observed by the SP (see Fig.\ref{fig:ZED}).

The ZED switches states to fingerprint the RSs perceived by the SP. It uses FSK modulation to generate an N-bit synchronization sequence, with each bit's duration $T^b=\frac{L}{2}TTI$, where $L$ is the number of RS symbols per bit, and $R^b=\frac{1}{T^b}$ is the bit rate. To transmit a bit 0 (or 1), the ZED generates a periodic signal at frequencies $F_0$ (or $F_1$), alternating between transparent and active modes \cite{AmB-FSK-Alto}. The frequencies are orthogonal, ensuring maximal detection probability.

In this paper, we generate the input signals using two Hadamard sequences, $c_0$ and $c_1$, each consisting of $N_c$ chips with chip time $T_c$, so that $T^b = N_c T_c$. Specifically, $x_0(t) = \sum_n rect_{T_c}(t - nT_c) \cdot c_0[k]$ and $x_1(t) = \sum_n rect_{T_c}(t - nT_c) \cdot c_1[k]$, where $c_i[k] \in \{0,1\}$. These sequences are orthogonal and balanced in terms of $0$ and $1$ counts. For example, the sequences $0 0 1 1 0 0 1 1$ and $0 0 0 0 1 1 1 1$ satisfy these conditions. Unlike the initial model, we define $T_c$ first and adjust the sequence length to optimize performance.

The synchronization signal is then generated with a pseudo-random bit sequence noted $b=[b_1 \dots b_{N_b}]$, of $N_b$ bits: 
\begin{equation}
\label{eq:refseq}
    x(t)=\sum_{n=0}^{N_b-1}b_n x_1(t-nT^b)+\sum_{n=0}^{N_b-1}(1-b_n)x_0(t-nT^b)
\end{equation}
where $b_n$ corresponds to the $n^{th}$ bit value. The numerical simulations of section IV are obtained with the sequence proposed in \cite{AmB-DemoLTE-Alto}, which is made of a combination of Barker codes $BC=[bc_+(7),bc_+(7),bc_-(7)]$, with $bc_+(7)$ the 7-length Barker code and $bc_-(7)$ the same flipped sequence.


Having defined the input signal, let us now focus on the signal measured at the SP. As in \cite{ZED-Loc-SY}, we denote $\mathbf{h}\in\mathbb{C}^K$ the vector of the propagation channel coefficients, where each element $h(k)$  corresponds to the channel coefficient of sub-carrier $k$. In this model, the channel coefficients remain constant during the synchronization sequence. 

Figure \ref{fig:ZED} illustrates the channel changes with the ZED mode. In transparent mode, only the BS-to-SP reference channel is active, i.e., $\mathbf{h}=\mathbf{\Gamma}$. In active mode, a secondary path from the ZED reflection is added, resulting in $\mathbf{h}=\mathbf{G}=\mathbf{\Gamma + \Phi.\Lambda}$. For simplicity, we assume no reflection in transparent mode and perfect reflection in active mode, but the key point is that the observed channel at the SP changes slightly. $\mathbf{\Gamma}, \mathbf{\Phi}, \mathbf{\Lambda} \in \mathbb{C}^{K \times 1}$, with $|\Gamma| >> |\Phi \Lambda|$. These values are constant over time but vary with frequency.

Then, using the propagation channel model and the signal generated by the ZED, the received signal by the SP at the $l^{th}$ RS on sub-carrier $k$ is $y(k,t(k,l))$, noted $y(k,l)$ for brevity:
\begin{equation}
y(k,l)=e^{j\phi(k,l)}\sqrt{P_u}\left( \mathbf{\Gamma} (k)+\mathbf{\Lambda}(k)\mathbf{\Phi}(k)x(k,l)\right)+\alpha(k,l)
\end{equation}
 $\alpha(k,l)\sim\mathcal{N}\mathbb{C}(0,\sigma^2)$ is a complex Gaussian noise with zero mean and variance $\sigma^2$, and  $\phi(k,t)$ stands for phase variations along successive TTI, due to the receiver oscillator. This later cannot be estimated and imposes the use of a non-coherent receiver.
 
 Note that the exact time $t(k,l)$ depends on the 4G/5G numerology but our notations above remain general. As illustrated in Fig.\ref{fig:LTE}, the $l^{th}$ RS symbols of all sub-carriers are not necessarily aligned.

\begin{figure}[htbp]
\centerline{\includegraphics[scale=0.6]{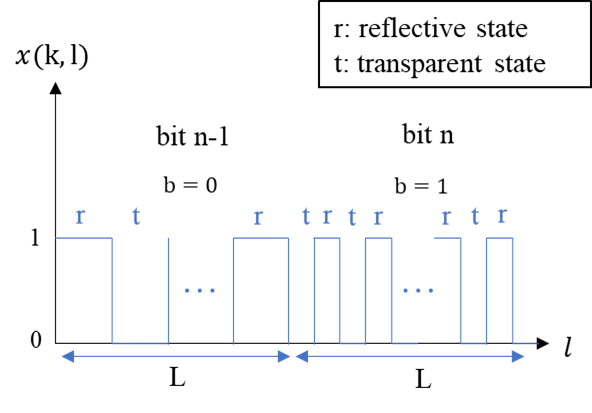}}
\caption{Signal generated by the ZED for two succesive bits with $b_{n-1}=0$, $b_{n}=1$. }
\label{fig:ZEDseq}
\end{figure}

\section{Optimal detector}
We propose an optimal ZED detector based on minimum mean square error (MMSE) path gain estimation followed by a Neyman-Pearson detector to minimize false alarms and maximize detection probability. In Section A, we derive a robust estimator using samples from a single sub-carrier, optimal when synchronized with the input sequence, and study its robustness to time shifts in Section B. Section C introduces a maximal ratio combiner for aggregating observations from multiple sub-carriers, which may have different OFDM symbol times. Finally, Section D presents a Neyman-Pearson test to determine the optimal threshold for a given false alarm probability.

\subsection{Single frequency receiver}
\begin{figure}[htbp]
\centerline{\includegraphics[scale=0.42]{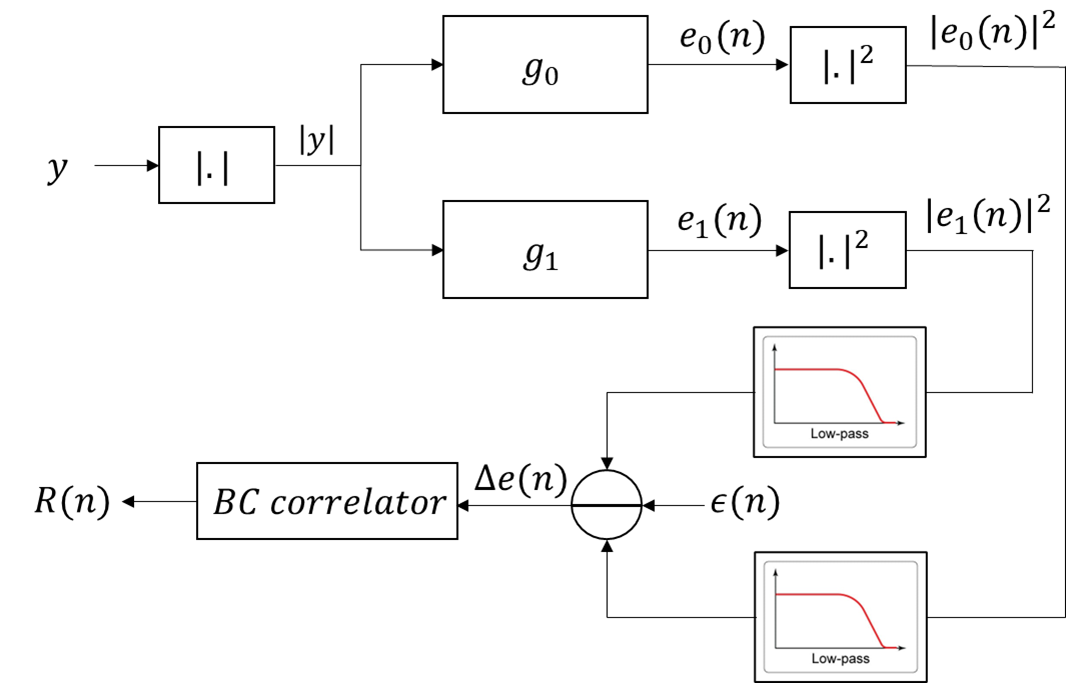}}
\caption{Optimal detector with the 2 correlators, and the contrast $R(n)$. The pseudo-filters $g_0$,$g_1$ are introduced in section III-A and the low pass filters in section III-B.}
\label{fig:detection_model}
\end{figure}

In this section only the indices $k$ are omitted for the sake of clarity since we work with a single sub-carrier.
A non coherent detector based on the absolute value of the received signal, to be robust against random phase noise at the receiver. 
The full structure of the proposed receiver is introduced in Fig.\ref{fig:detection_model}. 
Let us detail the expressions of inputs and outputs of this receiver. 
\begin{itemize}
    \item If ZED is in transparent mode:
\begin{equation}
\label{eq:transparent}
\begin{split}
|y(l)|=|\sqrt{P_u}\Gamma+\alpha(l)|\approx \sqrt{P_u} |\Gamma| + \alpha_r(l).
\end{split}
\end{equation}
\item If ZED is in reflective mode:
\begin{equation}
\label{eq:active}
\begin{split}
|y(l)|=&|\sqrt{P_u}(\Gamma+\Lambda\mathbf\Phi)+\alpha(l)|\\
& \approx \sqrt{P_u} \left(|\Gamma| + \gamma \right)+ \alpha_r(l),
\end{split}
\end{equation}
\end{itemize}
where $\gamma=real(\Lambda\Phi e^{-j\varphi(\Gamma)})$, (with $\varphi(\Gamma)$the phase of $\Gamma$) is the projection of $\Lambda\Phi$ on $\Gamma$ in the complex plane. It may be positive or negative.
The noise $\alpha_r(l)\sim\mathcal{N}(0,\frac{\sigma^2}{2})$ is real valued.
The former approximations are valid under the condition that $|\Gamma|>> |\Phi\Lambda|$ and $\sqrt{P_u}|\Gamma|>> |\alpha|$, which is reasonable in practical conditions. 

For each bit period $T^b$, the receiver aims at detecting if a bit 0 or a bit 1 is present, or nothing.



The optimal detector is built with two perfect correlators given by centered version of the pulsed shape input waves $g_0(t) = 2x_0(t)rect_{T_b}(t)-1$ 
and $g_1(t)=2x_1(t)rect_{T_b}(t)-1$, sampled at the position of the pilots. Then, the outputs are combined according to the pseudo-random sequence as illustrated in Fig.\ref{fig:detection_model}. 

Importantly, we use modified correlators to comply with missing RS data or non regular RS, and to resample the observation at the OFDM symbol rate, which will facilitate the combination of multiple sub-carriers observations. The correlators, at OFDM symbol $n$ (corresponding to time $t_n=n\cdot T^{ofdm})$ are given by:
\begin{equation}
\label{eq:EE}
\begin{split}
    e_0(n)&=\sum_{\tiny\begin{matrix}l; t(l)\in \\ [t_n,t_n+T^b]\end{matrix}}  \frac{y(l)}{a_{0}(n)}\cdot \mathds{1}_{x_0(t(l))} - \frac{y(l)}{b_{0}(n)}\cdot \mathds{1}_{\bar{x_0}(t(l))} \\
    e_1(n)&=\sum_{\tiny\begin{matrix}l; t(l)\in \\ [t_n,t_n+T^b]\end{matrix}}   \frac{y(l)}{a_{1}(n)}\cdot \mathds{1}_{x_1(t(l))} - \frac{y(l)}{b_{1}(n)}\cdot \mathds{1}_{\bar{x_1}(t(l))} 
    \end{split}    
\end{equation}
where $\mathds{1}_{c}$ is the indicator function that equals $1$ if $c$ is true, and $0$ otherwise. Also, $a_i(n)=\sum_{l; t(l)\in [t_n,t_n+T^b]}  \mathds{1}_{(g_i(t(l))=1}$ and $b_i(n)=\sum_{l; t(l)\in [t_n,t_n+T^b]}  \mathds{1}_{(g_i(t(l))=-1}$. $a_i(n)$ and $b_i(n)$ count respectively the number of observations under transparent or reflective modes respectively, and verify $a_i(n)+b_i(n)=L$. Their role in \eqref{eq:EE} is to normalize the values in transparent and reflective modes, to guarantee that the term $|\Gamma|$ in $y_l$ (\eqref{eq:transparent} and \eqref{eq:active}) is canceled, even if the sampling is irregular. It also permits to get the output $e_0(n)$ and $e_1(n)$ sampled at the desired rate. Efficient implementation of these pseudo-filters is feasible, with almost the same computational load as a conventional filter.

\begin{lem}
Given $x_0(t)$ and $x_1(t)$, the two aforementioned orthogonal sequences on a period $T^b$, and given that the ZED active bit is $i\in\{0,1\}$, with its complement noted $j=\bar{i}$,
the outputs of the two correlators are:
        \begin{equation}\label{eq:ei}
        e_i(n)= \sqrt{P_u} \gamma+\alpha_i(n)
        \end{equation}
        \begin{equation}\label{eq:ej}
        e_{j}(n)=\alpha_j(n)
        \end{equation}
        with $\alpha_i(n) \sim \mathcal{N}(0,\frac{\sigma^2}{2}(\frac{1}{a_i(n)}+\frac{1}{b_i(n)}))$ 
and $e_i(n), e_j(n) \in \mathbb{R}$.
\end{lem}
Note that in these expressions, it is assumed that $n$ is such that $n T^{ofdm} = t_0 + k T^b$, where $t_0$ is the starting time of the ZED sequence, and $k\in\mathbb{N}$.  
The asynchronous case is introduced in section III.B.

\begin{IEEEproof}
The proof is straightforward if the two sequences are orthogonal, by introducing \eqref{eq:transparent} and \eqref{eq:active} in \eqref{eq:EE}. The noise variance derives directly from the sums in \eqref{eq:EE}. Note that having $a_i(n)+b_i(n)=L$, the variance is minimal if $a_i(n)=b_i(n)$ which is granted with a regular sampling with no missing observations. In this case, a conventional correlator can be used.
\supprime{\paragraph{Proof of (\ref{eq:e0_0})}
         \begin{equation} 
            \begin{split}
            &|e_0(n_s)|\\ &=|\sum_{l=n_s}^{L+n_s}\sqrt{P_u} (|\Gamma| + real(\Lambda\Phi) x_0(l))g_0(l)+ \alpha_r(l)g_0(l)|\\
            &=|\sqrt{P_u}|\Gamma|\sum_{l=n_s}^{L+n_s} g_0(l)+ \sqrt{P_u} real(\Lambda\Phi)\sum_{l=n_s}^{L+n_s}x_0(l)g_0(l)\\
            &+\sum_{l=n_s}^{L+n_s}\alpha_r(l)g_0(l)
            \end{split}
         \end{equation}
        \SY{where $g_0$ is bipolar waves of +1/-1.Thus, in the first term, $\sum_{l=n_s}^{L+n_s}g_0(l)=0$. In the second term, the sum of $x_0g_0$ is equal to L/2 because with the alignment of the input sequence and the correlator, half of the terms are zero, thus $\sum_{l=n_s}^{L+n_s}x_0(l)g_0(l)=L/2$. Idem for $|e_1(n_s)|$ in (\ref{eq:e1_1}).}

\paragraph{Proof of (\ref{eq:e1_0}) }
\begin{equation} 
            \begin{split}
            &|e_1(n_s)|\\ &=|\sum_{l=n_s}^{L+n_s}\sqrt{P_u} (|\Gamma| + real(\Lambda\Phi) x_0(l))g_1(l)+ \alpha_r(l)g_1(l)|\\
            &=|\sqrt{P_u}|\Gamma|\sum_{l=n_s}^{L+n_s} g_1(l)+ \sqrt{P_u} real(\Lambda\Phi)\sum_{l=n_s}^{L+n_s}x_1(l)g_1(l)\\
            &+\sum_{l=n_s}^{L+n_s}\alpha_r(l)g_1(l)
            \end{split}
         \end{equation}
          \SY{The different in (\ref{eq:e0_0}) and (\ref{eq:e1_0}) comes from the fact that $g_1$ is proportional to $x_0$ while $g_0$ is orthogonal, and then the second term in the sum cancels. Idem for $|e_0(n_s)|$ in (\ref{eq:e0_1})}.}
\end{IEEEproof}
Based on these two outputs, the objective is now to determine if the ZED active bit is present in $e_i(n)$, while $e_j(n)$ is used as a reference measure of a noise only signal. The proposed receiver is derived in three steps: first, we propose an unbiased estimor of $\eta^2=P_u|\gamma|^2$, that characterizes the power of the ZED propagation path; second, we use the BC sequence (defined in \eqref{eq:refseq}) to optimally combine $N_b$ observations; third the estimations made over different sub-carriers will be combined in section III-C.

\begin{lem}
    Given the two observations $e_i(n)$ and $e_j(n)$, an unbiased estimator $\tilde{\eta}$ of the ZED path $\eta$ is:
    \begin{equation}\label{eq:eta_lem2}
        \tilde{\eta}(n)= |e_i(n)|^2 - |e_j(n)|^2 - \epsilon(n),
    \end{equation}
    with $\epsilon(n)=\sigma^2\left(\frac{1}{a_i(n)}+\frac{1}{b_i(n)} - \frac{1}{a_j(n)} - \frac{1}{b_j(n)}\right)$
\end{lem}
Note that if the sampling is regular and symmetric with $a_i=a_j=b_i=b_j=L/2$, then the correction term $\epsilon(n)$ cancels. 

\begin{IEEEproof}
    From \eqref{eq:ei} it comes out that $|e_i(n)| \sim \mathcal{R}\left(.;\eta,\frac{\sigma}{\sqrt{2}}\sqrt{\frac{1}{a_i(n)}+\frac{1}{b_i(n)}}\right)$, and $|e_j(n)| \sim \mathcal{R}\left(.;0,\frac{\sigma}{\sqrt{2}}\sqrt{\frac{1}{a_j(n)}+\frac{1}{b_j(n)}}\right)$,
    where $\mathcal{R}\left(.;\nu,\tau\right)$ stands for the Rice distribution with center shift $\nu$ and scale $\tau$.
    Then, knowing that a Rice distributed r.v. $z$ verifies
   $\mathbb{E}[|z|^2]=\nu^2+2\tau^2$, one obtains:
   \begin{equation}\label{eq:ei2}
       \mathbb{E}[|e_i(n)|^2]= \eta^2 + \sigma^2\left(\frac{1}{a_i(n)}+\frac{1}{b_i(n)}\right),
   \end{equation}
   and
    \begin{equation}\label{eq:ej2}
        \mathbb{E}[|e_j(n)|^2]= \sigma^2\left(\frac{1}{a_j(n)}+\frac{1}{b_j(n)}\right),
    \end{equation}
    According to \cite{talukdar1991estimation}, this estimator corresponds to the maximum likelihood (ML) estimator and is unbiased.
\end{IEEEproof}
Then, the pseudo-random bit sequence used to generate $x(t)$ is now used as a correlator leading to:
\begin{equation}\label{eq:Rn}
    R(n) = \frac{1}{N_b} \sum_{m=0}^{N_b-1} (2b_{m} -1) \cdot \Delta e(n+m\delta t) - \epsilon(n+m\delta t)
\end{equation}
with $\delta t=\frac{T^b}{T^{ofdm}}$.
When $n_0$ is such that $n_0T^{ofdm}=t_0$, then~:
\begin{equation}\label{eq:Rn2}
\begin{split}
    R(n_0) &=  \frac{1}{N_b} \sum_{m=0}^{N_b-1} e_{b_m}(n_0+m\delta t) ^2 - e_{1-b_m}(n_0+m\delta t)^2\\
    &-\epsilon(n_0+m\delta t)\underset{N_b\rightarrow \infty}{\rightarrow } \eta^2 
\end{split}
\end{equation}
Invoking the central limit theorem, when $N_b$ is large enough, $R(n_0)$ can be approximated by a Normal r.v.: 
\begin{equation}
\label{eq:RwithZED}
    R(n_0) \sim \mathcal{N}(\eta^2,\sigma^2 \cdot \lambda_{n_0}),
\end{equation}
with $\lambda_{n_0}=\frac{1}{N_b}\left(\frac{1}{a_i(n_0)}+\frac{1}{b_i(n_0)} + \frac{1}{a_j(n_0)} + \frac{1}{b_j(n_0)}\right)\geq \frac{4}{L N_b}$.
\supprime{The BC correlator $BC_{corr}$ in fig \ref{fig:detection_model} is defined as
\begin{equation}\label{eq:bccoor}
            BC_{corr}= \frac{1}{N_b}\left\{ \begin{array}{rcl}
            1 & \mbox{for}
            & bit=1 \\ -1 & \mbox{for} & bit=0 
\end{array}\right.
\end{equation}
}
Note that if the ZED is absent, the signal contains noise only, and for any $n$:
\begin{equation}
\label{eq:RwithoutZED}
    R(n) \sim \mathcal{N}(0,\sigma^2 \cdot \lambda_{n}),
\end{equation}

\subsection{Synchronization and time shift}
The expression in \eqref{eq:Rn2} is valid only for a perfect synchronization $n=n_0$. In order to exploit this sequence for synchronization purpose, we need to guarantee a good peak to side-lobe ratio, when the receiver observes the input signal, for any $n$.

Let first consider a bit-time shift, meaning that the delay is given by $n=n_0+k\cdot \delta t$. In this case, \eqref{eq:Rn2} is modified according to the auto-correlation of the bit sequence. Given $\rho_b(m)=\frac{\Phi_b(m)}{\Phi_b(0)}$ the normalized  autocorrelation of the symmetric ($+1/-1$) bit-sequence used by the ZED, one obtains from \eqref{eq:Rn}: 
\begin{equation}\label{eq:Rn3}
\begin{split}
    R(n_0+k\delta t) \underset{N_b\rightarrow \infty}{\rightarrow }  \eta^2\cdot\rho_b(k)
\end{split}
\end{equation}
The amplitude of these peaks is thus controlled by the bit-sequence autocorrelation. Barker-codes sequences are known to be optimal for short-length codes. 

Now we can also consider an additional shift of $q \delta t_c$, with $\delta t_c=\frac{T_c}{T^{ofdm}}$. Considering the sequences $c_0$ and $c_1$, $e_i(n_0+k\delta t+ qT_c)$ decreases while $e_j(n_0+k\delta t+ qT_c)$ increases. For additional delay below $T_c$, one obtain smooth variations of the contrast value. 

To reduce the sensitivity of the detector to time errors (e.g. small time jitter due to the low-cost ZED), we introduce a low pass filter with a cutting frequency higher than $1/T_c$. This reduces the time accuracy of the time detection, but this is not an issue since the primary objective is to detect the ZED, not to synchronize on the system.  This is illustrated in section IV. For the sake of visibility, the filter is introduced before the BC correlator (see Fig.\ref{fig:detection_model}). To conclude the width of the main peak is in the order of $T_c$, and the peak to side lobe gain directly depends on the bit sequence, with 
\begin{equation}
    G_{peak}=20\log10(\rho_M),
\end{equation}
with $\rho_M=max_m \rho_b(m)$.

\supprime{
(\ref{eq:R}) and (\ref{eq:yes seq}) are modified as a function of the time shift $\tau$, depending on the auto-correlation function of the input sequence. Thus: 
\begin{equation}
       R(\tau) = HN_s\sqrt{P_u}real(\Lambda\Phi)\cdot \left( G_{00}(\tau) - G_{01}\right)(\tau) +v,
\end{equation}
where $G_{00}(\tau)$ and $G_{01}(\tau)$ are the autocorrelation and intercorrelation functions of $g_0(t)$ and $g_1(t)$, respectively.
These sentences have to be designed such that $G_{00}(\tau)<< 1; \forall |\tau|>\epsilon$ and $G_{01}(\tau)<< 1; \forall |\tau|>\epsilon$, for some small $\epsilon$.
The optimization of these random sequences is kept out of the scope of this paper, but follows standard developments in coded transmission such as for WCDMA. The current version we use is a combination of 7-length Barker code (BC) sequences, one of these being inversed. We focus in the next section of the False Alarm/Detection probabilities for a given sequences. 
}

\subsection{Multiple frequency receiver and Hypothesis Test}
In this section we focus on the synchronous case, i.e. where $n=n_0$. 
In the previous subsection, we derived an optimal estimator of the ZED path $\tilde\eta^2$, that we will now note $\tilde\eta^2(k)$ for $k^{th}$ sub-carrier. $\sigma^2$ is independent of the sub-carrier $k$. Indeed, given that the environment is multi-path, the true parameter $\eta^2$ is frequency dependent: $\eta^2(k)$.

Based on equation \eqref{eq:Rn}, denoting $R(n_0,k)$ for the $k^{th}$ sub-carrier estimation, we now develop an optimal detector (in the Bayesian sense) \cite{neyman1933,tse2005} combining multi-frequency observations, assuming that the theoretical ZED path, $\eta^2$, is constant with $k$, for the sake of simplicity.

The hypothesis test corresponding to a given sequence is:
\begin{itemize}
    \item \Hyp{0}: The searched sequence is not present, and the contrast is just random due to the noise. $R(n_0,k)$ is normally distributed as given by \eqref{eq:RwithoutZED}.
    \item $\Hyp{1|\eta^2}$: The ZED is present, with a given ZED path $\eta^2$.
$R(n_0,k)$ is distributed according to \eqref{eq:RwithZED}.
\end{itemize}
\begin{figure}[htbp]
\centerline{\includegraphics[scale=0.45]{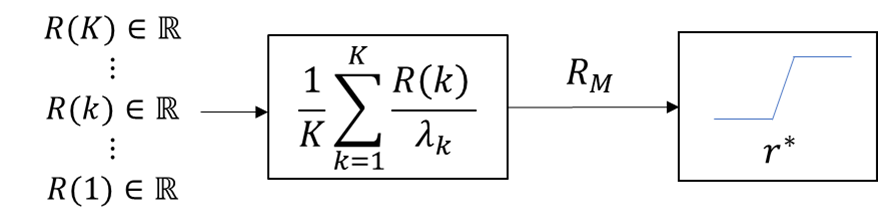}}
\caption{Optimal detector based on multiple subcarriers.}
\label{fig:comb}
\end{figure}

\begin{lem}\label{lem:Htest}
Given the hypothesis test above, any optimal decision in the Bayesian sense is given by 
\begin{equation}
R_M \testH r^*,
\end{equation}
where $R_M=\frac{1}{K}\sum^{K}_{k=1} \frac{R(n_0,k)}{\lambda_{n_0}(k)}$ is the average of the likelihood factor over all the subcarriers.
\ $r^*$ is the decision threshold. This detector is represented in Fig.\ref{fig:comb}.
\end{lem}

\begin{IEEEproof}
The proof relies on the well-known detection theory \cite{poor2013} stating that any optimal test can be expressed as a log-likelihood ratio test, given by:
\begin{equation}
\label{eq:testh}
\Lambda(\boldsymbol{R})=\log\left(f(\boldsymbol{R}|\Hyp{1})\right)-\log\left(f(\boldsymbol{R}|\Hyp{0})\right) \testH r,
\end{equation}
where:
\begin{equation}
\begin{split}
    \log (f(\boldsymbol{R}|\Hyp{0})) &=- \frac{1}{2}\sum_k\log(2\pi\sigma^2 \lambda_{n_0}(k))\\
    &-\frac{\sum_k R^2(n_0,k)/\lambda_{n_0}(k)}{2\sigma^2},
    \end{split}
\end{equation}
and 
\begin{equation}
\begin{split}
     \log (f(\boldsymbol{R}|\Hyp{1|\eta^2})) 
     &=- \frac{1}{2}\sum_k\log( 2\pi\sigma^2\lambda_{n_0}(k))\\
     &-\frac{\sum_k |R(n_0,k)-\eta^2|^2/\lambda_{n_0}(k)}{2\sigma^2}.
\end{split}
\end{equation}
Combining these two log-likelihood in \eqref{eq:testh} leads to: 
\begin{equation}
\label{eq:testh2}
\Lambda(\boldsymbol{R}) =  \frac{2\eta^2\sum_k(R(n_0,k)/\lambda_{n_0}(k)) - \eta^4\sum_k\frac{1}{\lambda_{n_0}(k)}}{2\sigma^2},
\end{equation}
and keeping on the left side only the terms with $R$:
\begin{equation}
\sum_k \frac{R(n_0,k)}{\lambda_{n_0}(k)} \testH r\cdot f(\eta^2),
\end{equation}
where $f(\eta^2)$ is a function that inversely proportional to $\eta^2$, which means that the test threshold $r^*$ depends on the value of the hidden parameter $\eta^2$. 
\end{IEEEproof}

\subsection{Neyman Pearson test}

Defining the threshold $r^*$ would need to know prior information relative to the test: the respective costs of false alarm and non-detection, and the prior probability of \Hyp{0} and \Hyp{1}, and also the value of the unknonw coefficient $\eta^2$ correspoding to the ZED. Instead, we propose to use the Neyman-Pearson approach \cite{poor2013}, that allows to predefine a target probability of false alarm $p^*_{fa}$. This probability is easily defined from the application perspective: at which rate do we accept to get wrong ZED detection? The optimal threshold $r^*$ can be then estimated, and the detection probability $p^*_D(\eta^2)$ can be evaluated as a function of the ZED path strength. 
The application of this approach to ambient backscatter was first introduced in \cite{AmB-detection-zargari2024}, but the pilot signal and noise channels differ from those in our paper, and they did not account for the contrast in their model.
\begin{lem}
    The false alarm probability $p_{fa}$ associated to the detector of Lemma\ref{lem:Htest} is given by:
\begin{equation}
\label{eq:pfa}
    p_{fa}=\mathbb{P}_{R_M|\Hyp{0}}[r>r^*]=Q(\frac{r^*}{\sqrt{var}})
\end{equation}
where $var=\frac{\sigma^2}{K}\sum_k \lambda_k$.

The threshold $r^*$ corresponding to the target $p^*_{fa}$ is: 
\begin{equation}
\label{eq:seuil}
    r^*=\sqrt{var}\times Q^{-1}(p^*_{fa})
\end{equation}

Last, the detection probability can be deduced as:
\begin{equation}
\label{eq:pd}
    p_D(\eta^2)=\mathbb{P}_{R_M|\Hyp{1}}[r>r^*]=Q(\frac{r^*-\eta^2}{\sqrt{var}})
\end{equation}
\end{lem}
\begin{IEEEproof}
$p_{fa}$ is the probability that $R_M$, given \Hyp{0}, is above the threshold $r^*$. Under \Hyp{0}, $R_M$ is the average of $K$ random variables of variance $\sigma^2\lambda_k$, leading to:
\begin{equation}
    R_M^{H0}\sim\mathcal{N}(0,\frac{1}{K}\sigma^2\lambda_k).
\end{equation}
Based on this distribution, \eqref{eq:pfa} is obtained, with $Q(.)$, the tail distribution function of the standard normal distribution. 
Under $\Hyp{1|\eta^2}$, $R_M$ is given by
\begin{equation}
\begin{split}
        R_M^{H1}&=\frac{1}{K}\sum^{K}_{k=1}\tilde\eta^2=\eta^2 + z,
\end{split}
\end{equation}
where $z\sim \mathcal{N}(0,\frac{1}{K}\sigma^2\lambda_k)$, which allows to obtain the lemma.
\end{IEEEproof}


\section{Performance Evaluation by Experiment}
The experiments presented in this work were conducted using the FIT/CorteXlab testbed \cite{massouri2014cortexlab}. This testbed, capable of supporting up to 40 nodes, offers a highly versatile platform that enables a wide range of experimental scenarios. For the purposes of this study, we utilized a two-node configuration: one node acting as the transmitter and the other as the receiver.
\subsection{Experiment setups}
\begin{figure}[htbp]
\centerline{\includegraphics[scale=0.28]{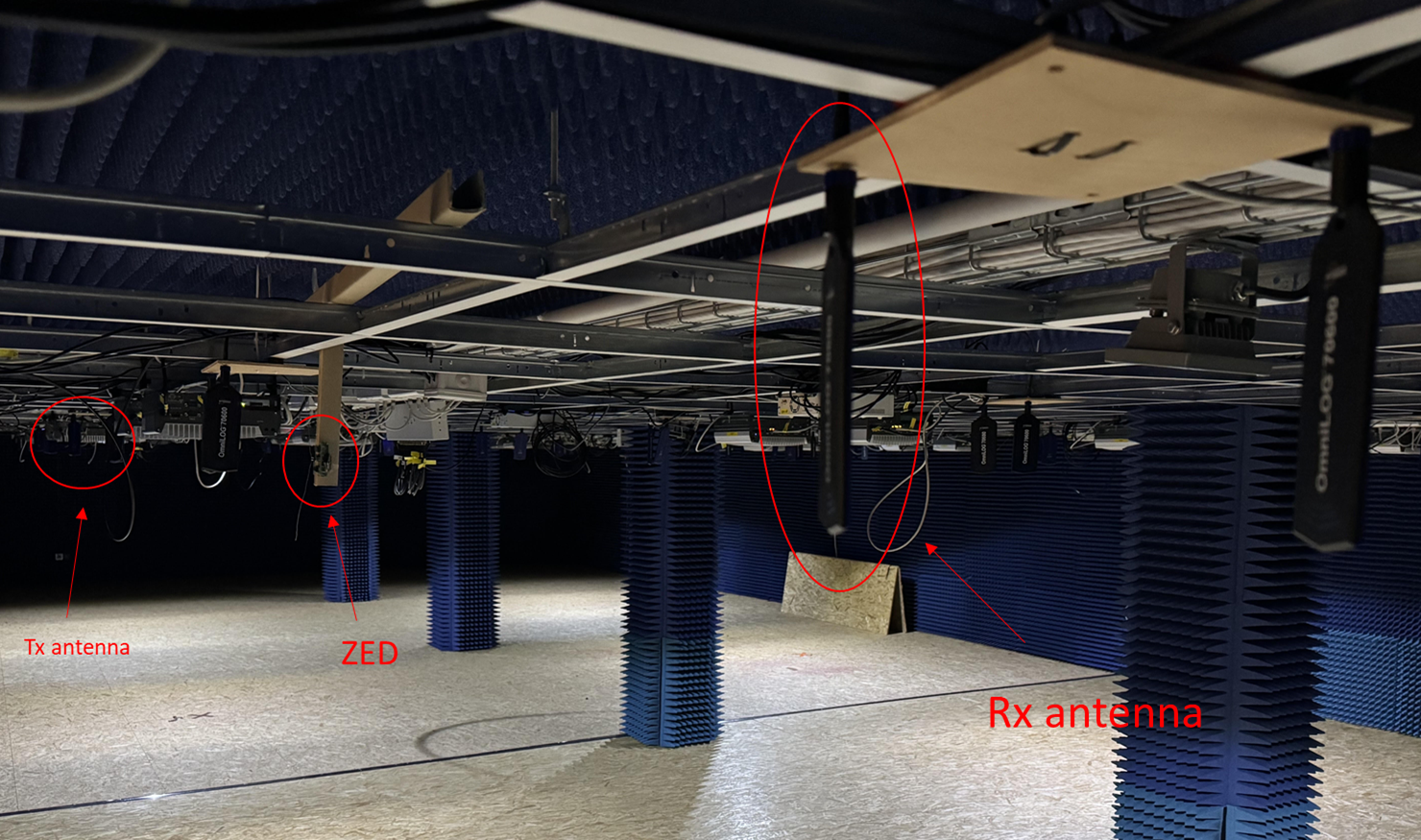}}
\caption{experimental environment in CortexLab}
\label{res:CortexLab}
\end{figure}

The experiments were conducted in a controlled environment (Figure \ref{res:CortexLab}) to evaluate the detection mechanism's performance by measuring the peak-to-lobe ratio. LTE parameters were used: each frame consists of 14 OFDM symbols, 2 of which are reference signals for the ZED signal. The experiment had a bandwidth of $B=2.5$ MHz, an FFT length of 128 ($FFT_{len}$), with 8 subcarriers as the cyclic prefix (CP). The OFDM symbol time, including the CP, was $T^{ofdm}=57.6 \, \mu s$. FSK modulation frequencies for bit 0 and bit 1 were $f_0=125$ Hz and $f_1=500$ Hz, respectively. A 4th-order Butterworth low-pass filter with a cutoff frequency of $f_{cf}=100$ Hz reduced the detector's sensitivity to time errors (Figure \ref{fig:detection_model}).

\subsection{Simulation results}
\begin{figure}[htbp]
\centerline{\includegraphics[scale=0.33]{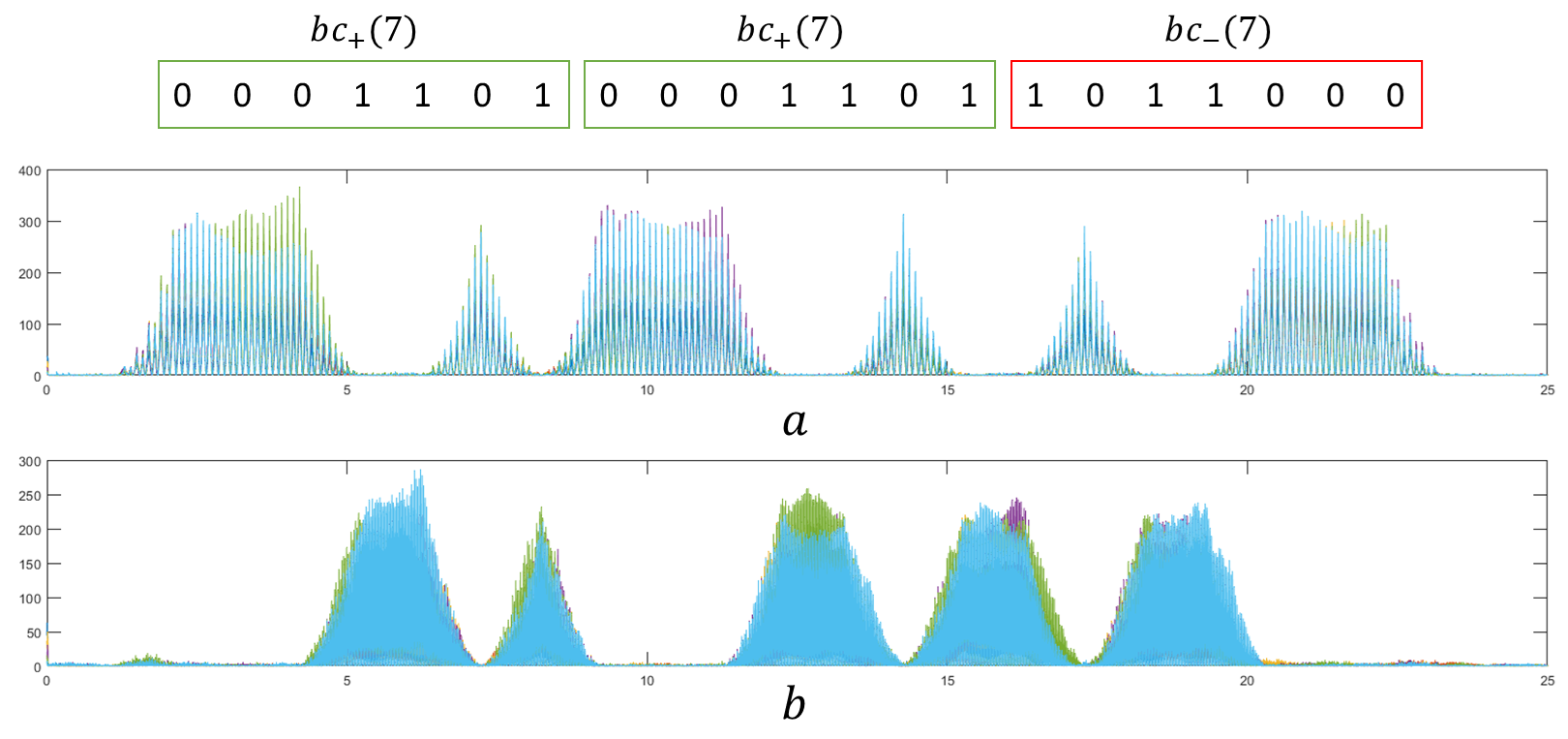}}
\caption{Experimental results with 6 different sub-carriers: $|e_0|^2$ in (a)  and $|e_1|^2$ in (b) corresponding resp. to bits 0 and 1.}
\label{res:e}
\end{figure}

Figure \ref{res:e} shows the outputs of \( |e_0|^2 \) and \( |e_1|^2 \) for the BC sequence \( bc_+(7) = [0001101] \) across 6 sub-carriers, each in a different color. The x-axis represents the number of bits, and the y-axis represents amplitude. The matched filter in Figure \ref{fig:detection_model} produces staggered outputs, alternating up and down in line with the BC code.

\begin{figure}[htbp]
\centerline{\includegraphics[scale=0.35]{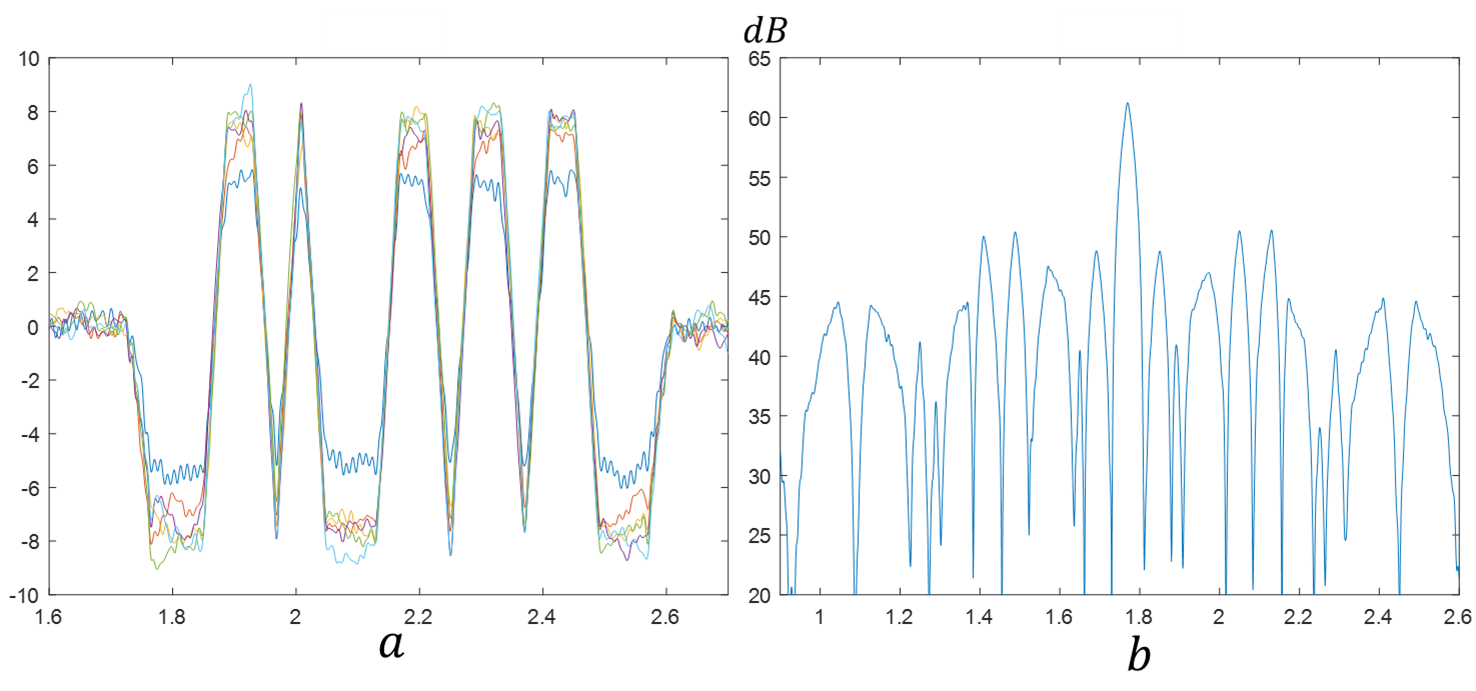}}
\caption{experimental results of: (a) output of the contrast $\Delta e(n)$; (b) output of the combiner $R_M$ (in dB scale)}
\label{res:R}
\end{figure}

Figure \ref{res:R}(a) displays the contrast \( \Delta e \) for the 6 sub-carriers, with the x-axis in seconds and the y-axis in amplitude. The final result, \( R_M \), after the BC correlator and combiner (Figure \ref{fig:detection_model} and Figure \ref{fig:comb}), is shown on a logarithmic scale in Figure \ref{res:R}(b). The peak-to-lobe ratio is about \( 11 \, \text{dB} \), confirming the effectiveness of our detection method.


\section{Conclusion}
In this paper, a new detector for the ZED based localization system studied in \cite{ZED-Loc-SY} is proposed. The solution is based on a correlator, a combiner, and a Neyman-Pearson optimal detector is introduced. Experimental results on the FIT/CorteXlab testbed demonstrated the model’s effectiveness, achieving a peak-to-lobe ratio of 11 dB. The use of BC sequences improved detection reliability, and the results confirm the model’s accuracy even in noisy conditions. Future work will focus on optimizing the detection for multiple ZEDs, and the choice of code to improve the peak-to-lobe ratio. 


\section*{Acknowledgement}
This work is partly supported by the European Project Hexa-X II under (grant 101095759) and by BPI France under the program France Relance (5G Events Labs).

\supprime{}
\bibliographystyle{ieeetr}  
\bibliography{reference}   






\end{document}